\documentclass[fleqn,twoside]{article}
\usepackage{espcrc2}

\usepackage{graphicx}
\usepackage[figuresright]{rotating}

\newcommand{\AmS}{{\protect\the\textfont2
  A\kern-.1667em\lower.5ex\hbox{M}\kern-.125emS}}

\title{A snap shot from the history of cosmic ray research:\\
                   a Japanese scientist view}

\author{Yasushi Muraki\address[DPKU]{Department of Physics, Konan University  
        Kobe 658-8501, Japan}%
        \thanks{Correspondence email address:muraki@stelab.nagoya-u.ac.jp, muraki@konan-u.ac.jp }}

\begin{document}

\begin{abstract}
This short note describes the history of cosmic ray research.
A part of this document was presented orally at the international conference 
of CRIS 2010 held in Catania, Italy.  The document
is written being based on the English translation of a Japanese article entitled 
"One Hundred Years of Research on Cosmic Rays".  
The document was published in 2008 by the Japan Astronomy Society 
as a series of books on "Modern Astronomy" (Volume 17). 
\vspace{1pc}
\end{abstract}

\maketitle

\section{Introduction}

 First of all, I should like to thank the organizing committee for having given me an opportunity to talk 
about the history of cosmic ray research.  It is, of course, impossible to give a comprehensive survey of 
one hundred years of cosmic ray research in a short time.  So I must apologize at the outset that 
this will be just a snapshot from our long history that I have heard from pioneers.  
Recently, I had two opportunities to review the history of cosmic ray research.  
I published an illustrated narrative on "What Are Cosmic Rays?"for younger readers [1], 
and also an article entitled "One Hundred Years of Research on Cosmic Rays".  
The latter was published by the Japan Astronomy Society in a series of books on Modern Astronomy that 
were written in Japanese [2].  For this reason, and because next year will be the centenary of the discovery 
of cosmic rays, it seemed an appropriate time to present my personal views 
on the history of cosmic ray research in English.  
The early history was published before in 1985 in a book edited by Sekido and Elliot [3]. 
However, we note this did not include all contributions.

\section{Back to early Days (1900-1911)}
The start of cosmic ray research was deeply connected with the two famous discoveries: 
X-rays by Roentgen and radio-activity by Becquerel.  These were discovered both in year of 1900.  
The discovery of cosmic rays 
was made when people were pushing forward the study of radio-activity.  The mineral discovered by 
Madame Curie and Schmidt in 1896 called pitchblende was an indispensable key mineral for the study 
of the properties of the radio-activity, since they showed extremely strong radio-activity.  When people 
tried to put the stone including the mineral on the table and measured its intensity 
as a function of the distance, 
people noticed a curious fact that the flux did not go to zero even at the time 
in which the observer measured 
the intensity from a long distance from the sample.  What are the components 
that they could not remove 
from the base line of the data? Theodor Wulf carried the electrometer on the Eiffel tower, 
but the meter did not show zero as expected.

\section{Early Days of Cosmic Ray Research (1911-1928)}

Three scientists challenged this matter. Hess (Austrian), Kolhoerster (German) and Gockel (Swiss).  
They used the balloon flight to study the difference of the ionization as a function of the altitude.  
They got on the balloon and courageously flied up over 5,300m height, carrying an instrument called the electrometer.  
They might have been very cold at the highest point and might have strong headache due to low atmosphere pressure.  
However, they repeated the measurement of the ionization.  And finally, they found that the ionization increased 
when the balloon climbed up higher altitude. Hess named these radio-activities "high altitude rays" (Hoehen Strahlung).

   Who gave the name "cosmic rays" to these high altitude rays?  It was made by a famous US scientist, Millikan, 
who found the unit of the electric charge.  He carried the electroscope in the Rocky Mountain and sank the detector 
in Muir Lake (3,600m), located near the highest peak of the Rocky Mountain, Mt.Whitney (4418m) (Figure 1). 
Then he sank the detector into another lake of lower altitude (2000m) and compared the difference of the ionization.  
Then he found the difference purely came from just the absorption by the atmosphere and concluded that those activities were induced 
by the "rays" coming from the top of the Earth, i.e., the universe.  So he named it as "cosmic rays". 
It was the year of 1925. In his paper to the American Academy, he wrote that the penetrating rays were of cosmic origin 
and also he described that he sank the detector in a beautiful snow-fed lake.  This might be pointed out as the first experiment 
using the water of the lake as the detector located at high altitude.   It may be worthwhile to note that Millikan had a picture 
on the early universe that the early universe was made by a big nucleus.  So it would be natural to image 
that high-altitude rays are coming from large blob of nucleus [4].

\section{Study of Elementary Particles by means of Cosmic Rays (1932- present)}

   In the long history of cosmic ray research, many new particles have been discovered.   The first great discovery could 
be pointed out as the discovery of positrons, an anti-matter, which was made by Anderson in 1932.  It would be interesting 
to know that Millikan prepared the budget for the experiment of Anderson.  Then muons were found by Anderson 
and Nedermyer in 1937.  When muons were found, Yukawa believed that these particles might be the particle 
as he predicted in order to combine protons and neutrons inside nucleus to keep the nucleus stable (1935).  
However, the feature was different from his prediction. The meson theory once encountered a difficulty.  
In 1942, Sakata, Inoue and Tanigawa pointed out that the mesons found in cosmic rays might be of different kind 
from Yukawa mesons.  This hypothesis is called as "two kinds of meson" theory. In 1947, Lattes, Occhialini 
and Powell discovered the Yukawa meson in the emulsions exposed at high altitude weather station located 
at Mt. Chacaltaya in Bolivia (5,250m).  Thus the Yukawa meson theory was confirmed. The Nobel Prize 
was given to Yukawa in 1948 and to Powell in 1950.

    After that, a search for new particles in cosmic rays was steadily made.  However, until 1970, no new particle 
was discovered in cosmic rays and the job to find out new particles was considered as the matter 
of particle accelerator experiments.  Breaking this long silence, in 1970, short life particles were found 
in the emulsion experiments exposed at high altitude by the balloon flight by Niu et al [5].  
By present terminology, they are called as the charmed particles. A number of charmed particles 
were produced by the accelerators at BNL and SLAC in 1974 and their properties were investigated in detail.

    I must describe here also the quark hunting experiments extended around 1966, when I was 
a post-graduate course student of Nagoya University.  Around that time, scientists considered 
that quarks might be involved in cosmic rays. They might be produced by the nuclear interaction processes 
between very high energy cosmic rays and the atmospheric nucleus.  However, even though an enthusiastic 
search for the quarks has been made, they have not been found at all.  Finally, around 1974, a theory 
was proposed that quarks must be confined in a bag, called MIT bag model.  This arises from a special property 
between two quarks that have a quantum number "color".   Nowadays it is known as the "asymptotic freedom".  
The force between two quarks behaves completely in different way in comparison with the Coulomb force 
or the gravitational force.  When two quarks approach in a short distance, the force between two quarks 
becomes weak.  On the other hand, when two quarks separate each other in a certain distance, then strong 
attractive force appears between them.  Therefore, quarks are difficult to escape from the bag.   This concept 
is well known as "the confinement of the quark" in the bag.  In my opinion, this peculiar property between 
quarks has not been completely understood yet and the property must be classified into the same category 
of the physics: why there are heavy electrons (muons) in nature, why there are two kinds of particles 
that obey the Bose-Einstein statistics and the Fermi-Dirac statistics etc.  We must accept the concept as they are.

   Around 1980, a fantastic theory was proposed. Not only two forces, the electro-magnetic force and the weak force, 
but also three forces involving the strong force were unified. In other words, those three forces stem from the same origin. 
As a consequence of the theory, protons must decay into positrons and neutral pions. 
The theory is called as the Grand Unified Theory (GUT). If the prediction is correct, protons are expected 
to be no more stable. People constructed large-volume water-tanks in the underground mines and waited 
for the proton decay. However, what they saw was not the evidence of the proton decay but the neutrino 
burst produced by the Super Nova explosion appeared near the Large Magellanic Cloud in 1987. Koshiba 
got Nobel Prize thanks to this discovery. In addition, with the use of the same detector, another important 
evidence of the neutrino oscillation was found. Even now in the underground laboratories, many people 
are searching the evidence for the dark matter, WIMPs (Weakly Interacting Massive Particles).

\section{Study of the Universe by means of Cosmic Rays (1941- present)}

The study of the universe by means of cosmic rays may be alternatively defined as follows: where and how 
cosmic rays are accelerated. In other words, we are studying particle acceleration processes in various parts 
of the universe. According to the efforts extended by a number of scientists, we have now a good knowledge 
on the energy spectrum of cosmic rays. The energy range is extended over 12 orders of magnitude 
from $10^{8}$eV to$10^{20}$eV.
The intensity drops off quickly with the energy and typical flux 
at 1 GeV ($10^{9}$eV) is 100 particles/$m^{2}$/sec/str. 
They are sometimes produced in association with solar flares. 
By solar flares, particles are accelerated beyond 100 GeV. 
Forbush noticed at first that those high energy cosmic rays are coming from the Sun. 
They were observed in association with 
the large solar flare in February 28, 1942. 
According to our recent observation in association with the solar flare, particles 
are accelerated over 56 GeV [6]. 
 
 The flux of cosmic rays near 100 TeV ($10^{14}$eV) is approximately 1 particle/$km^{2}$/sec/str. They may be produced 
by the shock acceleration processes. A very prominent particle acceleration theory was proposed in 1977 
by five scientists [7], [8], [9]. It is now well known and called as the shock acceleration theory. 
Since these particles have been accelerated to such high energies, it is impossible to measure their energies 
by the magnet-spectrometer. Therefore, they are typically measured by the air shower method.
One of the clear evidence for the shock acceleration mechanism was obtained 
by the simultaneous observations in the X-ray region (ASCA) and Very High Energy gamma-ray region (HESS) (Figures 2 and 3). 
 
The number of sources observed in the VHE gamma-ray regions has been accumulated and it turns out to be higher than 50. 
In the near future, a powerful Cherenkov telescope CTA will be constructed. Then the number of sources will exceed 1,000. 
This means a new stage of cosmic ray research and we are really entering into the gamma-ray astronomy. 
In the new stage of VHE gamma-ray study, at first, people will classify those sources into proton acceleration sources 
and electron acceleration sources. Then within them we will search different sources 
that were produced by the different acceleration mechanism.

 About the diffuse gamma-rays sources, recently the study entered into a new stage.   Due to new discoveries 
by the surface detectors, MILAGRO and Tibet AS array, and also by the satellite board detector, 
FERMI gamma-ray telescope, many bright sources that emit gamma-rays have been observed. 
Also a global trend of the intensity difference (anisotropy) has been observed, which is induced by 
the modulation of the solar magnetic field (solar wind).   The MILAGRO team and 
Tibet team are going to construct new detectors in their observatories independently: 
one is at high altitude in Mexico (5,000 m) 
and the other is at high altitude in Tibet (4,200m). They are based on the water Cherenkov method and they 
are designed to be able to separate the showers originated by protons from gamma-rays.  In the latter case 
the muon components involved in the showers must be 50 times less than the showers induced by protons. 
We could find gamma-ray sources with energies higher than 100TeV by means of these experiments. 
Here, I would like to point out the possibility that the majority of cosmic rays might be accelerated 
by the stellar flares up to 100 GeV with energy spectrum of $E^{-3.7}$ 
(the first step acceleration) and those seed cosmic rays 
will be re-accelerated to 100 TeV 
by the shock acceleration process in the SNR with energy spectrum of $E^{-2.7} $[10].

 The cosmic rays beyond $10^{19}$eV, called the highest energy cosmic rays, are certainly the energy frontier. 
Accelerators could not produce such a high energy at the moment.   
The highest energy region is very attractive region 
for both particle physics and astrophysics. Protons with energy higher than $4\times10^{19}$eV 
could not be involved in our galaxy, 
so that their origin must be different from our Galaxy: either coming from other celestial objects or 
from the decay of exotic particles such as cosmic strings. However, at the moment, 
the observation results seem to include the systematic errors as shown in Figure 4.  
Probably the difference of the energy spectrum originates from the systematic errors 
associated with the determination of the energy of each shower. 
In order to determine the incident energy of the showers, 
results of the Monte Carlo calculation have to be considered. 
In order to obtain the correct energy spectrum, 
the Monte Carlo code of the simulation must be calibrated. However, there is no accelerator data 
on the forward production region of secondary particles beyond $10^{14}$eV. The LHCf experiment 
will soon provide important data for the calibration of MC code for each air shower group 
and we will be able to understand the correct feature of the nature around the GZK cut-off region. 
After that, as the next step of the research, large air shower experiments like EUSO and/or 
Auger North projects will be made. Then we may enter a new step on cosmic ray research 
with the energy higher than $10^{21}$ eV.

\section{A brief final story}
In closing this talk, I would like to introduce an interesting story that I heard from Prof. Oda, 
the father of X-ray astronomy of Japan together with Prof. S. Hayakawa. Since Oda stayed at MIT under Prof. Rossi 
when he was young, he could know why and how the air shower experiment started. Around 1951-1956, 
the existence of the magnetic field in the Galactic arm was found. Enrico Fermi proposed to 
Bruno Rossi to measure the cosmic rays around $10^{15}$eV by means of an air shower experiment. 
Since the strength of the magnetic field in the Galactic arm was already estimated to be 
about 3$\mu$gauss at that time, 
protons with energy higher than $10^{15}$eV could not be present in the Galactic arm, 
therefore a cut-off at $10^{15}$eV 
would have been expected. However, there is a shoulder near $10^{15}$eV but no cut-off for cosmic rays 
and the cosmic ray spectrum is extended over $10^{19}$eV. Anyway, 
this is a reason how the air shower experiment started at MIT.

\subsection{An additional story}
Finally, I would like to introduce another topics related to Japanese cosmic ray research.  
As I have already stated above section, 
around 1947, the arrival direction of cosmic rays had not known at all.  In 1950 Prof. Yataro Sekido 
intended to measure it.   In 1955, he planned to build a large Cherenkov telescope, with a diameter of 4m  [11].  
He planned to find sources of cosmic-rays to measure the intensity of cosmic-ray muons 
as a function of arrival direction in the sky.  People believed that cosmic rays arrived 
rectilinearly from their sources.  Around that time, the existence of the magnetic field 
in the Galactic arm was not known. 
 
    However, almost at the final stage of the construction, the existence of a weak magnetic field 
was discovered.  So Prof. Satio Hayakawa recommended to Sekido to stop constructing the telescope. 
However, Yataro Sekido insisted on finishing its construction and making observations. 
Hayakawa agreed with this policy.  
 Yataro Sekido actually made a sky map of muon intensity with the large telescope.  
He obtained an almost uniform arrival map of muons over the sky.  Then Prof. Nagashima 
was invited to Nagoya University to study the modulation of arrival directions of high energy 
cosmic-rays (about 10 TeV) by magnetic fields - either the solar magnetic field, 
the interplanetary magnetic field or the galactic magnetic field.  He found an anisotropy 
in arrival directions with an anisotropy of $\pm$ 2 $\times$ $10^{-3}$ over the mean intensity.  
Recently, with the use of an air shower detector located at the South Pole, a nice sky map 
of the anisotropy of cosmic-ray intensities at high energies was obtained.  Quite surprisingly, 
measurements of the Southern sky coincide with measurements of the Northern sky (Figure 5)[12].  
What I should like to say here is that the detection of cosmic-ray sources at the highest energies 
will be subject to the problems that occurred at lower energies.  So we should exercise 
some caution following Sekido's historical experience.

\begin{figure}[htb]
\includegraphics[width=5cm]{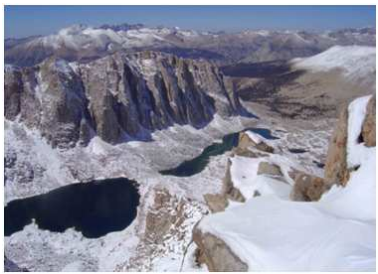}
\caption{The lake where Prof. Millikan sank the electrometer.}
\label{Fig1.eps}
\end{figure}

\begin{figure}[htb]
\includegraphics[width=7.0cm]{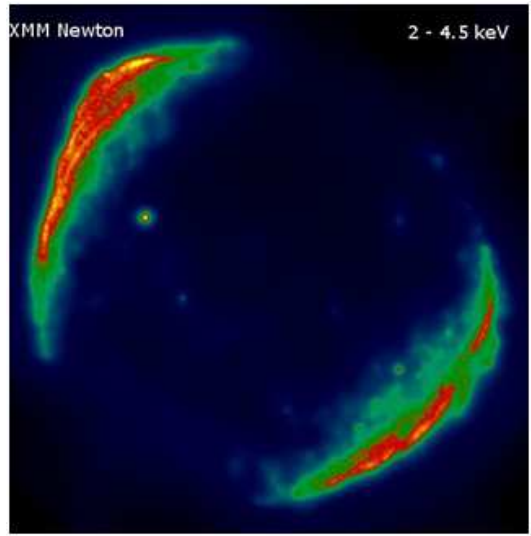}
\caption{The SN1006 observed by the X-ray telescope.}
\label{Fig2.eps}
\end{figure}

\begin{figure}[htb]
\includegraphics[width=7.0cm]{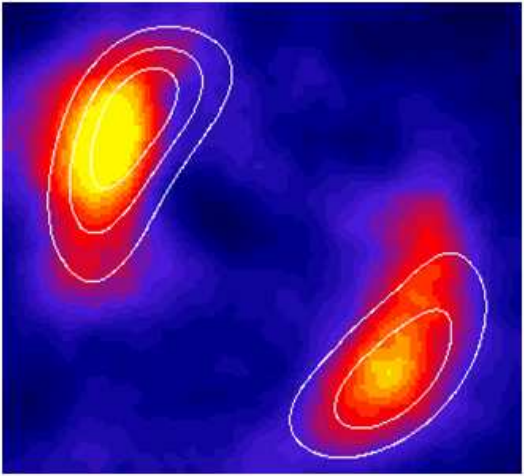}
\caption{The SN1006 observed by the HESS gamma-ray detector.}
\label{Fig3.eps}
\end{figure}

\begin{figure}[htb]
\includegraphics[width=7.0cm]{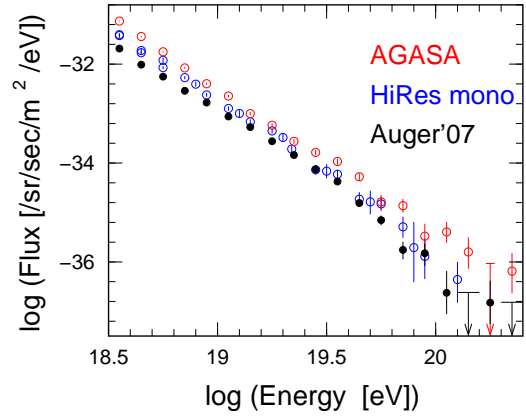}
\caption{The energy spectra observed by AGASA, Auger and Hi-Res.}
\label{Fig4.eps}
\end{figure}

\begin{figure}[htb]
\includegraphics[width=7.0cm]{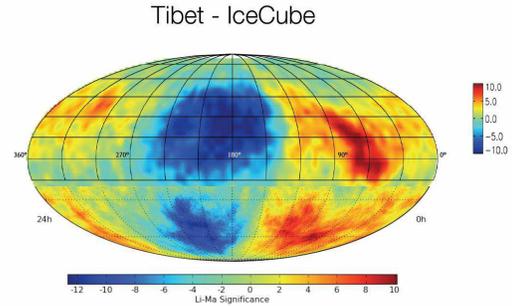}
\caption{The anisotropy of cosmic rays observed by Tibet AS array, Milagro (Northern sky)
and Ice Cube (Southern sky).  The dark area corresponds to 0.2$\%$ deficit of
cosmic ray intensity around 10TeV. }
\label{Fig5.eps}
\end{figure}

\end{document}